\documentstyle[preprint,aps]{revtex}
\tighten
\begin{document}
\draft

\title{\bf
SHELL MODEL IN THE COMPLEX ENERGY PLANE AND TWO-PARTICLE RESONANCES}

\author  {\rm R. Id Betan $^{a,b)}$,
              R. J. Liotta $^{a)}$,
              N. Sandulescu $^{a,c)}$,
              T. Vertse $^{a,d}$}
\bigskip

\address {\rm
  $^{a)}$~  Royal Institute of Technology, Alba Nova,
 SE-10691, Stockholm, Sweden \\
  $^{b)}$~  Departamento de Fisica, FCEIA, UNR,
 Avenida Pellegrini 250, 2000 Rosario, Argentina\\
  $^{c)}$~ Institute of Physics and Nuclear Engineering,
P.O.Box MG-6, Bucharest-Magurele, Romania\\
  $^{d)}$~ Institute of Nuclear Research of the
Hungarian  Academy of Sciences,
H-4001 Debrecen, P.O. Box. 51, Hungary}

\maketitle
\begin{abstract}
An implementation of the shell-model to the
complex energy plane is presented. The representation used
in the method consists of bound single-particle states,
Gamow resonances and scattering waves on the complex
energy plane. Two-particle resonances
are evaluated and their structure in terms of the
single-particle degreees of freedom are analysed. It is found
that  two-particle resonances are mainly built upon
bound states and Gamow resonances, but the contribution of
the scattering states is also important.

\end{abstract}

\pacs{PACS number(s): 25.70.Ef,23.50.+z,25.60+v,21.60.Cs}

\section{Introduction}
\label{sec:intro}

The study of processes taken place in the continuum part of
nuclear spectra is a difficult undertaking. This can be seen
by the large number of methods that have been proposed to
describe the continuum. Besides the Feshbach theory \cite{fes}
and its many versions, there are many  methods where
the continuum is described by a finite set of positive energy states.
These states are usually obtained by expanding the solutions of the 
nuclear
Hamiltonian in a harmonic oscillator basis  or by solving the
eigenvalue problem with box boundary conditions.
There have been also many variations of these procedures, e. g.
by using a transformed harmonic oscillators basis
\cite{sto} or harmonic oscillators with different frequencies
\cite{del}. The common feature of these representations
is that they are constrained to real energy solutions, as required
by quantum mechanics. However, the widths corresponding to
decay processes were evaluated by means of outgoing waves
(and therefore complex energies)
already by Gamow in his seminal paper on
barrier penetration \cite{gam}. The Gamow
states provide a natural definition of resonant states
\cite{siegert}.

Berggren \cite{b68}, and shortly afterwards
Romo \cite{r68}, showed that the bound and Gamow states together with
a set of complex scattering states form a representation
which spanns the space of complex energies.
In this representation the scalar product (the metric)
is non-Hermitian and
the norm of the Gamow states is defined
by using a method introduced by Zel'dovich \cite{zel}.
Later on techniques based on the complex rotation of the radial
distance 
have been also used to regularise the Gamow states and the matrix 
elements
involving them \cite{gya}. 

Microscopical calculations based on a single-particle representation
consisting of bound states, Gamow resonances and complex energy
scattering states (Berggren representation) were proposed some
years ago \cite{bl,lio}. This representation was recently expanded to
study two-particle resonances \cite{rol,pol}. It may be
surprising to see that it took
such a long time for us (after the paper in Ref. \cite{lio}) to arrive
to the two-particle representation. One of the reason of this delay is 
that
only recently we
have managed to find a method which allows one to
isolate the physical two-particle resonant states
from the  continuous background,
 as it was schematically outlined in Ref. \cite{rol}.
The problem is that only a small fraction of the calculated
states are physically relevant, since most of them (if not all)
are either wide resonances or part
of the non-resonant continuum. It is therefore important to be able
to isolate the physically meaningful states from the rest of the
spectrum. In this paper we will show how to achieve this. 
We will also show how two-particle resonances are built from
the single-particle degrees of freedom determining the dynamics
of the system.
This is important since it is not clear how a two-particle resonance
is formed. For instance, one may wonder to  what extent  such a state is
built upon particles moving in resonant states as well as
in non-resonant continuum states. Intuitevely one would say that for
the physically relevant two-particle resonances
one of the two particles
is in a narrow Gamow state while the other moves in bound or quasi-bound
states. Wide resonances and the
non-resonant continuum would play only a minor role, as assumed in Refs. 
\cite{pol,gol,gal,osv}. We will show that
this intuitive assumption is not always
supported by proper calculations.

Actually the question of the importance of the non-resonant continuum in 
the
calculations was usually skiped in relation to
processes taken place in the continuum part of the spectrum,
particularly regarding
radioactive decay where measurable life times
correspond to very narrow resonances. Therefore the calculation of the
corresponding decay widths can
be performed by using bound representations and the continuum itself
can be totally ignored \cite{bls}. This approximation, which was 
followed
by nearly
all available calculations of cluster decay (including alpha particles
and recent calculations of proton decay \cite{son,fmf,sor,kru})
has been very succesful in explaining experimental data
\cite{prep}. This may have contributed to the lack of interest
in methods which considered the continuum explicitely.
But with the development
of experimental facilities one could measure partial decay widths
of neutrons from giant resonances and  the continuum had to be included
explicitely in the formalism \cite{vlm,col,rrpa}.

Even more important, the experimental discovery of halo nuclei
triggered a very fruitful theoretical activity which showed
that halos cannot be understood without taking into account
wide resonances
and other elements belonging to the continuum  \cite{ebh,mit}.
All these elements are automatically included in the
representation presented in this paper.
The method, which is a generalization of the shell model to the
complex energy plane, was briefly given in Ref.\cite{rol}. We
will present it in detail here clarifying what a two-particle
resonance is. We will also provide an insigth on the influence
of the continuum upon the formation of two-particle
resonances.

The formalism is in Sect. 2. In Sect. 3 are the
applications  and a summary and conclusions are in Sect. 4.

\section{Formalism}
\label{sec:form}

The Berggren (one-particle) representation to be used in
this work has been described before in a number of situations,
e. g. in Refs. \cite{b68,bl,lio,rrpa}.
We will give here only a brief summary of the formalism.

The regular solutions of the Schr\"odinger equation
with outgoing boundary conditions corresponding to
a particle moving in a
central potential provide the single-particle bound states and
the so-called Gamow resonances.
The study of this case led \cite{b68,r68}  to  the
introduction of expansions for the Green- and  $\delta$-functions
in  terms  of the poles of the Green-function  plus  an  integral
along a continuum path in the complex energy plane, i. e.

\begin{equation}\label{eq:del}
\delta(r-r^\prime)=\sum_n
 w_n(r)w_n(r^\prime) + \int_{L^+} dE u(r,E)u(r^\prime,E)
\end{equation}
where we choose the integration path  $L^+$
to lie in the fourth quadrant
of the complex E-plane\footnote{This corresponds to the
second, or so-called non-physical, Rieman sheet.},
as seen in Fig. \ref{cont1}.
The summation runs over all bound states and poles of the Green
function (Gamow resonances) enclosed by the real E-axis and
the contour $L^+$.
One can choose quite general forms for the contour, as can be seen in
Ref. \cite{bl}, but it has to go through the origin and finish at 
infinite
on the real energy axis. However, as in any shell-model calculation,
one cuts the energies at a certain maximum value, which in Fig.
\ref{cont1} corresponds to the point $D\equiv (d,0)$. The other
points defining the contour in that figure are, besides the origin,
$A\equiv (a,0), B\equiv (b_1,b_2)$ and $C\equiv (c,0)$.

In Eq. \ref{eq:del} the scattering
functions on the contour are denoted by $u(r,E)$
while the wave functions of the bound
single-particle states
and the Gamow resonances are denoted by
$w_n(r)$. These states
have been indicated by open circles in Fig. \ref{cont1}.

An important feature in Eq. (\ref{eq:del}) is that the scalar
product is defined as the integral of the  wave
function times itself, and not its complex conjugate.
This is in agreement with the Hilbert metric on the real energy
since for bound states or for
scattering states on the real E-axis one can choose the phases
such that the wave functions are real quantities.
The prolongation of the integrand to the complex energy plane
allows one to use the same form for the scalar product
everywhere. This metric (Berggren metric) produces
complex probabilities, as has been discussed in detail in
e. g. Refs. \cite{bls,vlm}. Here it is worthwhile to
point out that for narrow resonances such probabilities
become virtually real quantities.

The integral in Eq.(\ref{eq:del}) can be discretized such that

\begin{equation}\label{eq:dis}
\int_{L^+} dE u(r,E)u(r^\prime,E)=
\sum_p h_p u(r,E_p) u(r^\prime,E_p)
\end{equation}
where $E_p$ and $h_p$ are defined by the procedure one uses
to perform the integration. In the Gaussian method $E_p$
are the Gaussian points and
$h_p$ the corresponding weights.  
Therefore the orthonormal (in the Berggren metric)
basis vectors $\vert \varphi_j\rangle$ are given by
the set of bound and Gamow states, i . e.
$\langle r \vert \varphi_n\rangle=\{w_n(r,E_n)\}$ and the discretized
scattering states, i. e.
$\langle r \vert \varphi_p\rangle=\{\sqrt{h_p}u(r,E_p)\}$. This
defines the Berggren representation.

By using the Berggren representation one
readily gets the two-particle shell-model equations, i. e.,

\begin{equation}\label{eq:tda}
(\omega_\alpha-\epsilon_i-\epsilon_j)X(ij;\alpha)=\sum_{k\leq l}
 <\tilde{kl};\alpha|V|ij;\alpha> X(kl;\alpha)
\end{equation}
where $\alpha$ labels two-particle states and $i, j, k, l$
label single-particle states. Tilde denotes mirror states
\cite{b68} and the rest of the notation is standard.

As usual, the two-particle wavefunction is,

\begin{equation}\label{eq:wf}
|\alpha>=\sum_{i\leq j} X(ij;\alpha) (c^+_ic^+_j)_\alpha|0>
\end{equation}
where

\begin{equation}\label{eq:amp}
 X(ij;\alpha) =  <\alpha|(c^+_ic^+_j)_\alpha|0>/(1+\delta_{ij})^{1/2}
\end{equation}
The discretization of the contour in Fig. \ref{cont1} produces
a series of points in the one-particle energy plane, as shown in
Fig. \ref{cont2}. Each point in this figure represents a state of
our one-particle Berggren representation.
Therefore the energy of the two-particle basis vector
$(c^+_ic^+_j)_\alpha|0>$, i. e.
$\epsilon_i+\epsilon_j$, is
the  sum of the point $i$ in Fig. \ref{cont2} with
the point $j$. Allowing the indices $i$ and $j$ to run over all the
one-particle basis states, ordered such that $i\leq j$, one obtains
the energies of the two-particle basis states (zeroth order energy)
in the corresponding (two-particle) complex
energy plane as shown in Fig. \ref{region1}.
One sees in this figure that the whole complex energy plane of
interest is covered by zeroth order solutions and, therefore,
it would be difficult in this plane to find the physical
two-particle states. This problem becomes more acute as the
number of elements in the basis increases, that is as the number
of points in Fig. \ref{cont2} becomes larger, because
the two-particle physical states would then be
embeded in a dense number of states belonging to the continuum.

A way to avoid this problem is to choose a one-particle contour
which leaves a physically relevant
two-particle complex energy region free of
zeroth order states. There are many possible contours
that satisfies this condition.
In Fig. \ref{cont3} we show one of such
contours and the corresponding Gamow resonances enclosed by it.
Since the bound states have to be considered in any case,
irrespective of the contour,
we do not include them in the discussion here.

The contour has a rectangular form defined by the points
$P_0\equiv (0,0), P_1\equiv (a,0), P_2\equiv (a,-c),
P_3\equiv (b,-c), P_4\equiv (b,0)$ and $P_5\equiv (d,0)$.
As mentioned above, the number $d$ should be infinite. However,
if one chooses $d$ large enough its value does not influence
significantly the
calculated quantities of interest. In the cases to be studied
in the next Section we will choose $d$ fulfilling such a condition.
It is also worthwhile to mention that, due to the Gaussian integration
method, the point $(0,0)$ will not belong to the Berggren
representation. The lowest energy on the contour corresponds
to the first Gaussian point, i. e. $(E_1,0)$.

By summing orderly the points of this figure with themselves
one obtains the two-particle states
shown in Fig. \ref{region2}. One can see
that if $2a<b$ then there is a region in the two-particle
complex energy plane which is practically free from any
uncorrelated solution.
Choosing the real energies $a$ and $b$ in Fig. \ref{cont3}
conveniently one can study
two-particle resonances lying in any reasonable energy region. We will 
call this the "allowed" energy region.
In Fig. \ref{region2} all possible two-particle
energy points have been drawn. The allowed region occupies
a rather small portion of the figure and therefore it
appears somehow diffuse among all the points. For clarity
of presentation we show in Fig. \ref{allow} the allowed region
and its neighborhood only.

The allowed region can be determined by fulfilling some
physically meaningful requirements. That is,
the correlated states of interest are those which live
a time long enough before
decaying. One would thus be able to observe them directly or
through effects that they induce, for instance through
the formation of halos. As we will show below, such states are built
mainly by Gamow resonances. Therefore the contour
in Fig. \ref{cont3} should be chosen in such way as to enclose
those resonances. Moreover, the values of $a$, $b$ and
$c$ should produce an allowed region where the calculated
two-particle states may lie. Since the two-body
interaction is attractive one expects that the correlated
low energy resonances would lie below their zeroth order
positions in Fig. \ref{allow}. Therefore the value
of $2a$ should be as small as possible while $b$ should be large,
which are just the conditions needed to obtain the allowed region.

Once the Berggren single-particle representation has been chosen,
the two-particle representation is built as in the standard shell
model, i. e. as the tensorial product of the one-particle
representation with itself. This defines the two-particle
space. In this space there are in zeroth order
two-particle configurations containing Gamow resonances only,
which in Fig. \ref{allow} are indicated by open circles.
Besides these, one sees that inside the allowed
region there are also configurations, indicated by dots,
corresponding to one particle moving in a Gamow resonance
and the other in continuum
states. These configurations can be seen more clearly
in Fig. \ref{region2}, where the dots form two
recognizable rectangles. Each rectangle corresponds to the
geometrical sum of a Gamow resonance, represented by an open
circle in Fig. \ref{cont3}, and the scattering
functions on the contour, represented by the crosses
in that Figure. In addition to the two-particle states
involving Gamow resonances there are also configurations,
denoted by croses in Figure 5, in which both particles are
in single-particle scattering states.

One expects that physically relevant resonances are mostly
determined by two-particle configurations
in which both particles are in either bound or
Gamow states. Since these states are the ones expected from the
standard shell model, we will call them "resonant shell model" states.

In the applications shown in this article the central field determining
the single-particle basis
will be of a Woods-Saxon type with a spin-orbit term as in
Ref. \cite{tvg}. The
corresponding bound and Gamow functions as well as the scattering
states will be evaluated by using the computer codes
described in Refs. \cite{tvg,ixa}.

As a two-body interaction we will use a separable two-body
force given by the derivative of the Wood-Saxon potential.
This effective interaction has shown to be satisfactory
to describe processes in the continuum, e. g.  neutron decay
from giant resonances \cite{vlm}. However, our purpose here
is not to explain in detail physical processes in the continuum,
but rather to understand the role played by the various ingredients
entering into the calculations.
The great feature of the separable force is that it does not
require the diagonalization of matrices. This is important
in our case since the dimension of the two-particle basis may be very
large and we want to evaluate all
states in order to examine the distribution of the energies
of the resonant shell model and scattering states discussed above.

Our effective interaction matrix element then is \cite{lan},

\begin{mathletters}
\label{eq:fpq}
\begin{equation}
<\tilde{kl};\alpha|V|ij;\alpha>  =  -G_\alpha f_\alpha(kl)f_\alpha(ij)
\label{mlett:1}
\end{equation}
where
\begin{eqnarray}
f_\alpha(pq) = && {1+(-1)^{l_p+l_q-\lambda_\alpha}\over 2
\sqrt{4\pi}}
(-1)^{j_q+1/2} \hat\lambda_\alpha \hat j_p \hat j_q
<j_p1/2~\lambda_\alpha 0|~j_q-1/2>\nonumber\\
&&\times
\int_0^{\infty}dr \varphi_p(r) (r\partial U(r)/\partial r) \varphi_q(r)
\label{mlett:2}
\end{eqnarray}
\end{mathletters}
and $U(r)$ is the volume part of the Woods-Saxon potential
defining the central field.
The rest of the notation is standard.

From Eqs. (\ref{eq:tda}) and (\ref{mlett:1}) one obtains the
dispersion relation,
\begin{equation}\label{eq:disrel}
-{1\over G_\alpha} =  \sum_{i\leq j} {f^2_\alpha(ij)\over
\omega_\alpha-\epsilon_i-\epsilon_j}
\end{equation}
which allows one to evaluate the correlated energies
$\omega_\alpha$.

Note that in Eq. (\ref{mlett:2}) the
angular part of the matrix element was evaluated according to the 
Hilbert
metric, while in the radial part the Berggren metric is used.
Note also that in Eq. (\ref{eq:disrel}) the square of the matrix
element appears, and not the square of its absolute value.

In the case of a separable force the amplitude of the
wave function can be written as
\begin{equation}\label{eq:wfa}
 X(ij;\alpha) =  N_\alpha {f_\alpha(ij)\over
\omega_\alpha-\epsilon_i-\epsilon_j}
\end{equation}
where $N_\alpha$ can be determined by the normalization
condition, i. e.
\begin{equation}\label{eq:nor}
N_\alpha^{-2} = \sum_{i\leq j} ({f_\alpha(ij)\over
\omega_\alpha-\epsilon_i-\epsilon_j})^2
\end{equation}

We will determine the strength of the separable force, i. e.
$G_\alpha$, by using the usual procedure of fitting the energy of a
two-particle state $\alpha$. 
However, in the drip lines nuclei which we will analyse 
there is
not any experimental data yet and, therefore, we will assume that such
state lies at a certain reasonable energy. We will also vary the
value of $G_\alpha$ thus obtained within a reasonable range in order
to study the influence of the two-body interaction upon the calculated 
states.

\section{Applications}
\label{sec:appl}

We will apply the formalism discussed above for two cases, one 
corresponding to a nucleus close to the neutron drip line and the other
close to the proton drip line. In both nuclei we will analyse
two-particle states with angular momentum $\lambda=0$. All 
partial waves with $l\leq 10$ will be included to evaluate the 
scattering states in the Berggren representation.

\subsection{Two-neutron resonances}
\label{sec:v40}

In this section we will present calculations for two-neutron
states in the double closed shell nucleus $^{78}$Ni. The
Woods-Saxon parameters are indicated in Table \ref{wsp} and the
corresponding single-particle energies are in Table \ref{sps}.
These
single-particle
states are quite similar to the ones given by a Skyrme-HF
calculations \cite{nic}.

As seen in Table \ref{sps}, the shell $N=50$
is well defined, since there is a gap of
about 3.6 MeV between the lowest particle state, which here
is $1d_{5/2}$, and the highest hole state, i. e. $0g_{9/2}$.

We will also evaluate a case where no bound single-particle
states are present. For this, we reduced the value of the
depth of the Wodds-Saxon potential to $V_0=37 MeV$. The corresponding
single particle energies are also given in Table \ref{sps}.
 Note that even in this case
the shell $N=50$ is rather well defined.

\subsubsection{The Fermi level is bound}

We will first analyse the case where there are
bound single-particle states, i. e. the case
$WS1$ in Table \ref{sps}.

Already from the start one faces the problem of determining which
single-particle states are to be included. In a standard shell-model
calculation one would include the states corresponding to a major
shell only plus possibly an intruder state. In our case that would
be the shell $N$=4 and the intruder state $0h_{11/2}$. However we
have now the imaginary part of the energy to be taken into
account. This may produce states with relatively small real parts
of the energies
but very large (in absolute value) imaginary parts. Such states
would induce very wide two-particle resonances and their
inclusion would imply the use of contours embracing large
portions of the complex energy plane which one would not expect
to influence the narrow two-particle resonances of interest.
We therefore decided to include all single-particle
states up to a real energy of 6
MeV and imaginary parts down to -4 MeV. This correspond to all
states shown in Table \ref{sps}.

As mentioned above, to determine the strength of the separable force we
will follow the standard procedure of adjusting
$G_\alpha$ by fitting the energy of a two-particle
state, which usually is experimentally known. In our
case we will assume that such state, which
would be the ground state of $^{80}$Ni, exists below
twice the energy of the lowest single particle state,
i. e. below 2$\epsilon_{1d_{5/2}}$. This energy
gap, i. e. the correlation energy, is more than
1 MeV in well established normal nuclei, like
$^{208}$Pb (where it is 1.244 MeV) and $^{56}$Ni
(1.936 MeV). However in our case the bound states are
so few and so slightly bound that such high energy gaps
do not seem to be reasonable. Since there is not any experimental
data which could guide us, and since our intention is just
to see how the strength of the force affects the results,
we will vary  $G_{\alpha}$ from zero to a maximum value corresponding
to a gap of 2.527 MeV, i. e. for a $^{80}$Ni(gs) energy of -4.183 MeV.
The value of the strength corresponding to the Berggren basis 
described below is $G_{\alpha}$= 0.0028 MeV.

In the calculations to be presented here
we used a rectangular contour with the vertices as in Fig.
\ref{cont3} with a=0.5 MeV, b=9 MeV, c=-4 MeV and d=20 MeV.
We thus include in the Berggren basis
all the Gamow states shown in Table
\ref{sps} plus the bound states $1d_{5/2}$ and $2s_{1/2}$.
With this contour the allowed region comprises the two-particle
energy plane with complex energies $(E_r,E_i)$ such that
$1 MeV <E_r< 9 MeV$ and $-4 MeV <E_i< 0 MeV$.

As already mentioned above, we will use
a Gaussian method of integration
over the contour. We found that in order to obtain
convergence within six digits in the evaluated quantities,
one has to include
10 Gaussian points for each MeV on the lines of the contour
in Fig. \ref{cont3},
except for the last segment
(the one going from (b,0) to (d,0)) where 5 points for
each MeV is enough.
We arrive to this conclusion
by always choosing the contour such that the resonances lie at least
300 keV from the borders of the contour. The number of scattering
states thus included in the basis is $n_g$=225.
The convergence of the results as a function of $n_g$
as well as the influence of the continuum upon the calculated
states will be given below.

One can check the reliability of the results by performing a calculation 
over
the real energy axis only. The real (bound) energies thus
obtained, which are the standard shell-model results and
therefore can be called "exact", should coincide with those evaluated
by using any contour. Moreover, when evaluating the strength of
the force by given a real energy $\omega_\alpha$ in
Eq. (\ref{eq:disrel}) the value of
$G_\alpha$ thus obtained should,
independently of the contour that one uses,
be a real quantity. 
All these requirements are fulfilled
in our calculations.
This is important since it is a strong test of the reliability
of our computing codes
as well as a confirmation of the validity of the formalism.

In Fig. \ref{g40} we show the energies evaluated by using
different values of the strength $G_\alpha$.
The energies follow the pattern discussed
in the previous Section. One can thus identify the distinctive
straight lines corresponding to scattering configurations.
For all G-values these lines appear practically in the same
position.
One can also see that the two-particle resonances can
readily be distinguished from the scattering states by
just looking at the figure. This is possible due to the
presence of an allowed region which only contains the physical
states.

In the figure the physical states are labelled by their zeroth
order configurations. One sees that as the interaction increases
the real part of the energies behave in a standard shell-model
fashion. Thus, the states with largest degeneracy feels the
interaction the most. Moreover, the ground state departs from the rest
of the spectrum more and more as the interaction increases, also
in agreement with expectations. Guided by these well established
facts one may assume that narrow configurations, which due to
the centrifugal barrier are usually the ones with highest
degeneracy, would follow a similar pattern. That is, they would
be dominant in buiding up the physical (narrow)
two-particle resonances, as it was assumed in Refs.
\cite{pol,gol,gal,osv}. But Fig. \ref{g40} shows just the opposite.
As the interaction increases all resonances become narrower,
except the one corresponding to the configuration $(0h_{11/2})^2$,
which becomes wider. This unexpected feature is a consequence
of the Berggren metric, i. e. of the
non-Hermitian character of the Hamiltonian matrix in the
complex energy sector. Similar properties, like violations of
the non-crossing of levels rule, were found in the
one-particle case \cite{fml}.

The behaviour of the physical resonances in Fig. \ref{g40}
is rather involved. Thus the states labelled
$(1d_{3/2})^2$ become first
narrower as the interaction increases but at a certain point,
for $G_\alpha$ = 0.002 MeV in the figure, this tendency is reversed.
At the same time the increase of -$Im(E)$ desaccelerates around the
same value of $G_\alpha$ for the states $(0h_{11/2})^2$.

To understand the behaviour of the states $(0h_{11/2})^2$
we give in Table \ref{wfh11} the main components of the
corresponding wave functions. A feature to be noticed 
is that in no case are bound configurations relevant.
As the interaction
increases wide configurations become more and more important.
This explains why the state becomes wider. However, 
the bound configuration $(1d_{5/2})^2$ start to become relevant
at a large enough value of $G_\alpha$. Thus at $G_\alpha$ = 0.002
MeV that configuration contributes with a value of 0.14 to
the wave function. This becomes more important as $G_\alpha$
becomes larger, thus desaccelerating the increase of the widths.

On the other hand, the wave functions of the states $(1d_{3/2})^2$
presented in Table \ref{wfd3} show that narrow configurations
(more exactly, configurations narrower than the zeroth order one,
which here is $(1d_{3/2})^2$) contribute substantially to the
structure of the state as the interaction increases, particularly
the narrow states $(0h_{11/2})^2$ and $(0g_{7/2})^2$ and the bound
configuration $(1d_{5/2})^2$. This explains why the states
$(1d_{3/2})^2$ 
becomes narrower as a function of $G_\alpha$. But as $G_\alpha$
increases scattering configurations become important
and, as a result, the states become wider.
An interesting feature in this contex is the sudden appearence
of a large contribution (of a value (0.42,0.18)) from a scattering
configuration at $G_\alpha$ = 0.0024 MeV. This corresponds to
the configuration $|C> = |0d_{3/2} c_{3/2}>$, where $c_{3/2}$
is the scattering function at (0.385,0) MeV. This is a
Gaussian point on the first border of the contour.
The energy of the configuration $|C>$ is (in MeV) (1.325,-0.479)
+ (0.385,0) = (1.710,-0.479), which is very close to the energy of the
resonance, i. e. (1.665,-0.447).  This can even be inferred
from Fig. \ref{g40} where the down open triangle
for the case $(1d_{3/2})^2$ being discussed here
practically overlaps with our continuum configuration $|C>$.
Therefore, according to Eq. (\ref{eq:wfa}), the corresponding
wave function component is large.

The unexpected behaviour of the resonances discussed above
is representative
for all the others in Fig. \ref{g40}, while
the bound states behave in a standard shell-model fashion.
It is perhaps surprising that the first excited bound states
(labelled $(2s_{1/2})^2$ in the figure) do not show any
remarkable sensitivity to scattering states, although they
lie close to the continuum threshold. Indeed, the
wave functions corresponding to these states consist
mainly of the configurations $(1d_{5/2})^2$ and $(2s_{1/2})^2$
for all values of $G_\alpha$.

One important feature of the calculation is that the
energies corresponding to physical states
converge to their exact values relatively fast as a function of the
dimension of the basis. This we show for the
states $(1d_{3/2})^2$ and $(0h_{11/2})^2$ in Tables
\ref{wf3conv} and \ref{wf11conv} respectively.
To assess whether the strength of the interaction affects the
convergence we have chosen different values of $G_\alpha$.
We thus see that indeed the energy corresponding to
$G_\alpha$ = 0.0004 MeV, which is the smallest
G-value shown in those Tables, coincides within a few keV with the exact
result already for $n_g$ = 0. But as $G_\alpha$ is increased
that agreement deteriorates. Particularly inadequate are
the energies evaluated by using $n_g$ = 0 for the states
$(0h_{11/2})^2$ and $G_\alpha\geq$ 0.0020 MeV. Not only are the
real parts of those energies wrong by an ammount ranging from
1.2 Mev (for $G_\alpha$= 0.0020 MeV) to almost 2 MeV, but
also the imaginary parts are large and positive, which
does not make sense since it would correspond, e. g.,
to negative widths. 
This last feature does not appear for the states $(1d_{3/2})^2$.

One can understand the deterioration of the resonant shell model
results
(i .e. of neglecting the continuum by using $n_g$ =0) as the
strength increases by noticing that it is trough the interaction
that continuum configurations become relevant in the calculation.
This also explains why the results corresponding to $n_g$ = 0
for the states
$(1d_{3/2})^2$ are generally better than those corresponding
to $(0h_{11/2})^2$ since here the interaction is
stronger (due to the degeneracy) for a given value of $G_\alpha$.
But already with $n_g$ = 10 the
agreement between the exact results and the approximated ones
is reasonable in all these cases of physical states. Moreover,
for $n_g$ = 100 the exact results are reproduced within 6 digits.
This convergence is better than the one required to achieve a
similar agreement in general, for which one needs the value
$n_g$ = 225 used in our calculations, as mentioned above.

Finally it is worthwhile to point out that the presence of
scattering states lying nearby physical states do not affect
the convergence, as seen in e. g. Table \ref{wf3conv}
for the state $(1d_{3/2})^2$
with $G_\alpha$ = 0.0024 MeV (cf. Figure \ref{g40}).

\subsubsection{The Fermi level is unbound}

In this subsection we will analyse the case where
there is not any bound single-particle states, i. e. the case
$WS2$ in Table \ref{sps}.

Actually there is not any essential difference between this
case and the previous one since within
this formalism all states (including the continuum states)
are treated on the same footing, indepedently
of the location of the Fermi level.

The single-particle resonances are wider than before and therefore
we used here a different one-particle contour, namely
a=0.1 MeV, b=13 MeV, c=-6 MeV and
d= 26 MeV.

We present in Fig. \ref{g37} the evaluated states as a function
of the strength $G_\alpha$, which we allowed to vary within
the same range as in the previous case.
The straight lines discussed above appear
also in this case with the same characteristics as before.

Even the physical resonances present the same features as in
the previous subsection. In particular, the states labelled
$(0h_{11/2})^2$ interacts strongly with all the others thereby
becoming wider while all the other states become narrower.

However there is a new important feature in this case,
namely the development of a bound state
which is induced by the two-particle interaction, as shown by the
states $(1d_{5/2})^2$. To analyse the reason of this
behaviour we present in Table \ref{wfd5} the main components
of the corresponding wavefunctions.
As expected, one sees that when the interaction is weak the
state is built practically by the configuration $(1d_{5/2})^2$
only. As $G_\alpha$ increases the two-particle resonance
approaches the continuum threshold and scattering states
contribute substantially to the wave function.
Thus, the state under the column labelled $scat$ corresponding
to $G_\alpha$ = 0.0012 MeV is $1d_{5/2}c_{5/2}$, where
$c_{5/2}$ is a scattering d-wave at energy (0.089,0) MeV.
At $G_\alpha$ = 0.0020 MeV the resonance approaches
threshold even more and here the continuum itself
becomes important. Indeed, the large contribution under
the column $scat$ corresponds now to the configuration
$c_{1/2}c_{1/2}$, where $c_{1/2}$ is a scattering s-wave
with an energy (0.011,0) MeV, itself very close to threshold.
As the interaction increases even more the state becomes
bound and at $G_\alpha$ = 0.0028 MeV the scattering states
cease to be important. But the interaction is here strong enough
to mix up all the shell model configurations, showing the
importance of Gamow resonances in inducing bound states in nuclei
that lie far from the line of $\beta$ stability.

\subsection{Two-proton resonances.}
\label{sec:pp}

 Proton resonances are usually narrower than the corresponding neutron
 ones due to the Coulomb barrier. It is therefore often in this case
 that one study many-body systems including only the narrow Gamow
 resonances. In this section we analyse this approximation for
 the case of two protons outside the $^{100}$Sn core.
 The single-particle
 proton states  correspond to the major shell N=4, which is the
 same as in the previous subsections. The core mean field
 is described by a Wood-Saxon potential with the  parameters given in
 Table \ref{wsp}. These parameters were adjusted
to obtain the single-particle states shown in
Table \ref{spsp}, which agree with systematics in this region.
Notice that none of these single-particle states is bound.

As in the neutron case analysed above
we include in our single-particle representation even states which
belong to higher shells, namely the states $1f_{7/2}$ and
$0i_{13/2}$, because they are relatively narrow. We include these
high lying shells in order to assess whether they can be neglected,
as one does within the standard shell model.

We chose even here the rectangular contour of Fig.
\ref{cont3} with vertices defined by the values $a$ = 0.1 MeV,
$b$ = 19 MeV, $c$ = -1 MeV and $d$ = 26 MeV. This contour
encloses all the Gamow resonances of Table \ref{spsp}.
Choosing the Gaussian points as indicated above in order to obtain
a precision of six digits, the number of scattering states for
each partial wave turns out to be $n_g$ = 298.

With the single-particle (Berggren) representation thus established
we calculated the complex two-particle energies
by solving the dispersion relation (\ref{eq:disrel}). The
corresponding wave functions were evaluated by using
Eq. (\ref{eq:wfa}).

We used in our calculations of the two-proton states, which would be
resonances in $^{102}$Te, values of the strength $G_\alpha$ in a
range similar to that in the neutron cases analysed in the
previous subsections. The results of the calculation are
shown in Fig. \ref{pp}. The general trends in this figure
are similar to the ones already found for the two-neutron cases.

One notices that even in this case where
all resonances are very narrow,
the narrowest resonance in zeroth order becomes
wider as the interaction increases while all the other
become narrower.
This is specially remarkable for the state that at zeroth order
is $(2s_{1/2})^2$, since one does not expect that a state with
such low degeneracy would be important to build up low lying
resonances. To analyse these states we present in Table
\ref{pps1} the corresponding wave function amplitudes
for values of the strength $G_\alpha$ used in Fig. \ref{pp}. 
As expected according to what we found for the neutron cases
above, the reason why these states become narrower is that
narrow configurations play an important role as the
interaction increases. However this trend is not as specific
as before, when just the narrowest neutron configuration (which
was $(0h_{11/2})^2$) contributed
most to the narrowing wave function. The equivalent shell is now
$(0g_{7/2})^2$, which first (at low values of $G_\alpha$) is
important but then decreases as the strength
increases. Perhaps even more amazing is the behaviour of
the shell $(1d_{3/2})^2$ which first
increase in importance but suddenly, starting at $G_\alpha$ =
0.0010 MeV, decrease again. These features indicate again
that the behaviour of the wave functions in the Berggren space
can follow patterns which are unusual from a standard shell model
viewpoint.
The only configuration in Table \ref{pps1} which increases
continuously in absolute value
as $G_\alpha$ increases is $(1d_{5/2})^2$, which
is also very narrow and may explain why these two-proton
resonances become narrower.

The other notable states in Fig. \ref{pp} are those labelled
$(1d_{5/2})^2$, which are very narrow for all values of
$G_\alpha$ and which rapidly decrease in energy as $G_\alpha$
increases, as expected for a pairing (ground) excitation.
Eventually the state becomes bound for a value of the strength
large enough, which in the figure is between 0.0014
MeV and 0.0016 MeV. To study the changing structure of these
pairing states we show in Table \ref{ppd5} the corresponding
amplitudes as a function of $G_\alpha$. As expected from truly
pairing vibrations \cite{mh}, the number of equally important
configurations increases with the strength of the pairing
force. Moreover the real parts of the wave function components
(which actually are virtually real numbers)
carry the phase $(-1)^l$, where $l$
is the orbital angular momentum of the corresponding single-particle
states. In this subject of pairing vibrations the
results of the method presented here and those of the standard
shell-model coincide.

In this case of very narrow Gamow resonances one notices
that in the two-particle wave functions the scattering states do not
seem to play an important role (cf. Table \ref{wfd3}). To analyse
this point we present in Table \ref{snen} the
dependence of the calculated energies, also for the states
labelled $(1d_{5/2})^2$ in Fig. \ref{pp}, upon the number of
scattering states $n_g$ included in the Berggren basis.
The general features of the results in this Table do not differ
much from those found in Tables \ref{wf3conv} and \ref{wf11conv}.
That is, for small values of the strength $G_\alpha$ the evaluated
energies reach fast its exact value as $n_g$ increases. But
this convergence wanes as $G_\alpha$ increases. Thus the
energy evaluated by neglecting the scattering states agrees with
the exact results within a few keV for $G_\alpha$ = 0.0002 MeV
but disagrees strongly for $G_\alpha$ = 0.0016 MeV.

It is interesting to see whether the corresponding wave functions
converge as badly as the energies do for large values of the
strength. This we show in Table \ref{snwf}, where we use the
extreme case $G_\alpha$ = 0.0016 MeV. Perhaps surprisingly, one
sees that the main components of the
wave functions evaluated for $n_g$ = 0 agrees
within a few percent with the exact ones.
This shows that the use of only narrow Gamow resonances, neglecting
the continuum, as was done in Ref. \cite{osv}, may be appropriate
to evaluate wave functions although the energies thus obtained
are inadequate.

\section{Summary and conclusions}
\label{sec:conc}

In this paper we have presented a formalism to evaluate
microscopically two-particle resonances within
the Berggren representation. This consists of bound states,
Gamow resonances and an infinite (continuous) set of
complex scattering states
lying on a contour in the complex one-particle energy plane.
The Gamow states included in the representation are those
enclosed by the contour. The scattering states appear as an
integral over the contour. We discretized this integral
by using a Gaussian integration procedure. Therefore the
infinte set of scattering states becomes reduced to
the finite value $n_g$ of Gaussian points. Using this
finite Berggren basis we constructed the two-particle basis
set of states as the tensorial product of the one-particle
basis with itself, as in standard shell model calculations.
We have shown that using an arbitrary contour one may get a
two-particle basis with energies covering the whole two-particle
complex energy plane of interest. This would hinder the evaluation
of two-particle states since they would be
embedded in a continuos set of basis states. To avoid this
drawback we have shown that there exists a contour that leaves
a region in the two-particle complex energy plane free of basis
states. It is just in this region where the physically
relevant resonances lie. Using that contour we have evaluated
all two-particle resonances with a precision of six digits by
choosing $n_g$ with values between 150 and 300, depending upon
the case under study. But we have found that with $n_g\approx 10$
one obtains a precision of a few keV for the energies of the
relevant resonances, while the corresponding wave functions are
provided within a precision of a few percent by neglecting the
scattering states altogether, i. e. with $n_g $ = 0.

We have applied the formalism to study neutron excitations in
$^{78}$Ni and proton ones in $^{100}$Sn.
The single-particle states were provided by a Woods-Saxon
potential and we chose a separable force as the two-particle
effective interaction.

For the neutron case we analysed a case where the Fermi level
was bound and another one where it lied in the continuum. In
both cases wide resonances were included in the basis. For
the proton case the Fermi level also lied in the continuum but
here all Gamow resonances were narrow.
We have shown that the position of the
Fermi level is irrelevant, since all basis
states are treated on the same footing.

We have shown that the
states which in zeroth order consist of configurations containing
scattering states feel the interaction very weakly.
Instead, the physical states consist mainly
of configurations containing only
bound states and Gamow resonances. These configurations are the
ones expected from the shell model. Even in cases where no bound 
configurations are present, the two-body interaction 
may induce narrow resonances and bound two-particle states. 
We found that the narrowest
of those configurations in zeroth order 
become wider as the interaction increases. At the same time, all
the other states become narrower.
This unexpected result, which
is induced by the Berggren metric, shows that
physically relevant resonances, i. e. narrow ones, may be strongly
influenced by states lying deep in the continuum.
Although the wave functions of the physical two-particle
resonances are mainly built upon shell model configurations, the
corresponding energies are strongly influenced by scattering
states.

Finally, it is important to point out that the 
application of the method presented here shows that it is a
natural generalization of the shell model to the
complex energy plane. 

\acknowledgments

This work has been supported by FOMEC (Argentina), by
the Hungarian OTKA fund Nos. T26244, T37991 and T29003, by
the Swedish Foundation for International Cooperation
in Research and Higher Education (STINT) and by
the Swedish Institute.

\begin{figure}
\caption{One-particle complex energy plane. It is shown
the contour $L^+$ corresponding to the energy of
the scattering waves (full line) and
the Gamow resonances enclosed by the contour
(open circles) defining the one-particle Berggren representation.
The bound states, which enter in the representation independently
of the contour, are not shown.}
\label{cont1}
\end{figure}
\begin{figure}
\caption{Discretized contour and Gamow resonances
defining the one-particle Berggren representation.
The open circles indicate the energies of the
Gamow resonances while the crosses are the energies of the
scattering states..}
\label{cont2}
\end{figure}
\begin{figure}
\caption{Zeroth-order two-particle energy points obtained from the
one particle states in Fig. \protect\ref{cont2}. These points
define the two-particle Berggren representation. The open circles
correspond to the cases in which both particles occupy
Gamow states. The dots
are the energies in which one particle is in a Gamow state
while the other is in a scattering state. The crosses are the energies
corresponding to the cases in which both particles are in scattering
states.}
\label{region1}
\end{figure}
\begin{figure}
\caption{One-particle discretized contour which produces the 
two-particle
energy region free of zeroth order states shown in Fig.
\protect\ref{region2}. The open circles indicate the energies of the
Gamow resonances while the crosses are the energies of the
scattering states. Note that the point (0,0) does not belong
to the representation. The lowest energy corresponding to
scattering basis states lies at $(E_1,0)$.}
\label{cont3}
\end{figure}
\begin{figure}
\caption{Energies of the uncorrelated two-particle states obtained
from the one-particle energies of Fig.
\protect\ref{cont3}. The open circles correspond to the cases
in which both particles occupy Gamow states. The dots
are the energies in which one particle is in a Gamow state
while the other is in a scattering state. The crosses are the energies
corresponding to the cases in which both particles are in scattering
states. The allowed region is the one with real energy between
2a and b and with imaginary energy larger than -c. The
basis vector with lowest energy lies at $2E_1$.}
\label{region2}
\end{figure}
\begin{figure}
\caption{As Fig. \protect\ref{region2} but enlarged such that
only the allowed region and its neighborhood is included.}
\label{allow}
\end{figure}
\begin{figure}
\caption{Energies of the calculated two-particle states
as a function of the strength $G_\alpha$ ($\times 10^4$,
in MeV) for the case $WS1$ of Table \protect\ref{sps}.
Only the allowed region in the two-particle energy plane
is shown. The straigth lines formed by small dots correspond to
continuum configurations where one particle is in a shell-model
state and the other in a scattering state.  The crosses
correspond to continuum configuration where both particles are in
scattering states.
The labels of the curves followed by the physical two-particle
resonances indicate the corresponding zeroth order configurations.}
\label{g40}
\end{figure}
\begin{figure}
\caption{Energies of the calculated two-particle states
as a function of the strength $G_\alpha$ ($\times 10^4$,
in MeV) for the case $WS2$ of Table \protect\ref{sps}.
Only the allowed region in the two-particle energy plane
is shown.
The straigth lines consisting of small dots correspond to
continuum configurations where one particle is in a bound or
gamow state and the other in a scattering state.  The crosses
correspond to continuum configuration where both particles are in
scattering states.
The labels of the curves followed by the physical two-particle
resonances indicate the corresponding zeroth order configurations.}
\label{g37}
\end{figure}
\begin{figure}
\caption{Energies of the physical two-particle states calculated 
as a function of the strength $G_\alpha$ ($\times 10^4$,
in MeV) for the proton case of Table \protect\ref{spsp}.
All physical resonances lying up to an energy of 10 MeV are
shown. Notice the scale in the imaginary part of the
energy, which indicates that the widths of the physical
resonances are in all cases small.
The labels of the curves followed by the physical two-particle
resonances indicate the corresponding zeroth order configurations.
The dashed lines were drawn to guide the eye.}
\label{pp}
\end{figure}
\begin{table}
\caption{Values of the Woods-Saxon parameters used in the
calculations. The spin-orbit parameters $r_0^{so}$ and
$a^{so}$ coincide with the ones corresponding to the volume
part given in this Table. The Coulomb radius in the
proton cases is the same as $r_0$, i. e. $r_0^{Coul}=1.19 fm$.
The meaning of these parameters is as in Ref. \protect\cite{tvg}.
\label{wsp}}
\begin{tabular}{cccccccc}
Core&$V_0 (MeV)$&$r_0 (fm)$&$a (fm)$&$V_{so} (MeV)$\\
$^{78}$Ni (neutrons) & 40 & 1.27 & 0.67 & 21.43 \\
$^{100}$Sn (protons) & 58.5 & 1.19 & 0.75 & 15 \\
\end{tabular}
\end{table}
\begin{table}
\caption{Single-particle neutron states in $^{78}$Ni
evaluated with the Woods-Saxon
potential given in Table \protect\ref{wsp}.
The complex energies are in MeV.
The column labelled $WS1$ corresponds
to $V_0 = 40$ MeV while $WS2$ to $V_0=37$ MeV.
The hole states $0g_{9/2}$ are given to show the magnitude of the
gap corresponding to the magic number $N=50$.
\label{sps}}
\begin{tabular}{cccccccc}
state&WS1&WS2\\
$0g_{9/2}$&(-4.398,0)&$(-2.587,0)$\\
$1d_{5/2}$&(-0.800,0)&(0.294,-0.018)\\
$2s_{1/2}$&(-0.295,0)&$-----$\\
$1d_{3/2}$&(1.325,-0.479)&(1.905,-1.241)\\
$0h_{11/2}$&(3.296,-0.013)&(4.681,-0.069)\\
$1f_{7/2}$&(3.937,-1.796)&(4.455,-2.851)\\
$0g_{7/2}$&(4.200,-0.167)&(5.799,-0.506)\\
\end{tabular}
\end{table}
\begin{table}
\caption{Main components of the wave functions corresponding to the
state which in zeroth order is $(0h_{11/2})^2$ as a function of
$G_\alpha$ ($\times 10^4$, in MeV) for the case $WS1$
of Table \protect\ref{sps}.
The corresponding two-particle energy E (in MeV) is also given.
Only components which in absolute value are larger than 0.2 are given.
The basis states are ordered according to their widths. Thus
$(0h_{11/2})^2$ is the narrowest and $(1f_{7/2})^2$ the widest
configuration.
\label{wfh11}}
\begin{tabular}{cccccccc}
$G_\alpha$&$E$&$(0h_{11/2})^2$&$(0g_{7/2})^2$
&$(1d_{3/2})^2$&$(1f_{7/2})^2$\\
8 &(5.399,-0.136)&(0.96,0.00)&(-0.27,-0.02)&------&------\\
12&(4.784,-0.342)&(0.92,0.02)&(-0.33,-0.03)&------&------\\
16&(4.283,-0.664)&(0.86,0.05)&(-0.35,-0.05)&(0.36,-0.19)&------\\
20&(3.975,-0.990)&(0.77,0.04)&(-0.33,-0.06)&(0.52,-0.11)
&(0.24,0.03)\\
24&(3.780,-1.222)&(0.70,-0.01)&(-0.32,-0.03)&(0.60,-0.01)
&(0.24,0.01)\\
28&(3.628,-1.376)&(0.66,-0.06)&(-0.31,-0.01)&(0.66,0.06)
&(0.23,-0.01)\\
\end{tabular}
\end{table}
\begin{table}
\caption{As Table \protect\ref{wfh11} for the
state which in zeroth order is $(1d_{3/2})^2$.
The relevant basis states (i. e. with amplitudes larger than 0.2)
include now the bound configuration
$(1d_{5/2})^2$ and configurations consisting of scattering states,
of which we give only the largest component under the column
$scat$.
\label{wfd3}}
\begin{tabular}{cccccccc}
$G_\alpha$&$E$&$(1d_{5/2})^2$&$(0h_{11/2})^2$&$(0g_{7/2})^2$
&$(1d_{3/2})^2$&$(1f_{7/2})^2$&$scat$\\
8 &(2.604,-0.809)&------&------&------&(1.00,0.01)
&------&------\\
12&(2.541,-0.679)&------&------&------&(0.99,0.03)
&------&------\\
16&(2.378,-0.506)&------&(-0.33,0.15)&------&(0.94,0.07)
&------&------\\
20&(2.059,-0.401)&(-0.25,-0.02)&(-0.46,0.06)&(0.23,0.04)
&(0.88,0.05)&------&------\\
24&(1.665,-0.447)&(-0.31,-0.08)&(-0.48,-0.00)&(0.25,-0.00)
&(0.70,-0.05)&------&(0.42,0.18)\\
28&(1.235,-0.683)&(-0.46,-0.22)&(-0.63,-0.04)&(0.33,0.01)
&(0.74,-0.19)&(-0.24,0.06)&------\\
\end{tabular}
\end{table}
\begin{table}
\caption{Convergence of energies corresponding to the
states labelled $(1d_{3/2})^2$ in Fig. \protect\ref{g40} as a
function of the number of Gaussian points $n_g$.
The value $n_g$=0
corresponds to the case where only bound states and Gamow
resonances are included in the basis.
The columns are labelled by the strength $G_\alpha$
($\times 10^4$, in MeV).
\label{wf3conv}}
\begin{tabular}{cccccccc}
$n_g$&4&20&24&28\\
0&(2.640,-0.896)&(3.299,-0.607)&(3.275,-0.858)&(3.227,-0.975)\\
10&(2.63416,-0.89697)&(2.13004,-0.42801)&(1.79169,-0.50527)&
(1.360,-0.82097)\\
50&(2.63448,-0.89643)&(2.05694,-0.39779)&(1.70880,-0.43585)&
(1.24293,-0.68603)\\
100&(2.63349,-0.89643)&(2.05889,-0.40198)&(1.67618,-0.44027)&
(1.23509,-0.68299)\\
150&(2.63349,-0.89643)&(2.05889,-0.40198)&(1.67618,-0.44027)&
(1.23509,-0.68299)\\
\end{tabular}
\end{table}
\begin{table}
\caption{As Table  \protect\ref{wf3conv} for the states
$(0h_{11/2})^2$.
\label{wf11conv}}
\begin{tabular}{cccccccc}
$n_g$&4&20&24&28\\
0&(6.018,0.004)&(2.777,0.320)&(2.237,0.681)&(1.775,0.860)\\
10&(6.02747,-0.03903)&(3.95235,-0.95715)&(3.75050,-1.18517)&
(3.59543,-1.33198)\\
50&(6.02693,-0.03949)&(3.97506,-0.98989)&(3.77995,-1.22224)&
(3.62815,-1.37649)\\
100&(6.02693,-0.03949)&(3.97507,-0.98988)&(3.77989,-1.22213)&
(3.62798,-1.37641)\\
150&(6.02693,-0.03949)&(3.97507,-0.98988)&(3.77989,-1.22213)&
(3.62798,-1.37641)\\
\end{tabular}
\end{table}
\begin{table}
\caption{Main components of the wave functions corresponding to the
state which in zeroth order is $(1d_{5/2})^2$ as a function of
$G_\alpha$ ($\times 10^4$, in MeV) for the case $WS2$
of Table \protect\ref{sps}. Under the column
$scat$ we give the largest component corresponding to
configurations consisting of scattering states.
The two-particle energy E (in MeV) is also given.
Only components which in absolute value are larger than 0.2 are given.
\label{wfd5}}
\begin{tabular}{cccccccc}
$G_\alpha$&$E$&$(1d_{5/2})^2$&$(0h_{11/2})^2$&$(0g_{7/2})^2$
&$(1d_{3/2})^2$&$scat$\\
4 &(0.538,-0.024)&(1.00,0.00)&------&------&------&------\\
12&(0.377,-0.010)&(1.00,-0.04)&------&------
&------&(0.21,0.23)\\
20&0.002,-0.000&(0.81,-0.04)&------&------
&------&(0.44,0.01)\\
28&(-0.700,0)&(0.84,-0.04)&(-0.32,-0.01)&(0.21,-0.01)
&(0.24,-0.11)&------\\
\end{tabular}
\end{table}
\begin{table}
\caption{Single-particle proton states in $^{100}$Sn
evaluated with the Woods-Saxon
potential given in Table
\protect\ref{wsp}. The complex energies are in MeV.
\label{spsp}}
\begin{tabular}{cccccccc}
state&Energy\\
$1d_{5/2}$&(2.583,-0.000)\\
$2s_{1/2}$&(4.007,-0.004)\\
$0g_{7/2}$&(4.469,-0.000)\\
$1d_{3/2}$&(4.917,-0.004)\\
$0h_{11/2}$&(7.559,-0.001)\\
$1f_{7/2}$&(9.710,-0.424)\\
$0i_{13/2}$&(16.361,-0.210)\\
\end{tabular}
\end{table}
\begin{table}
\caption{Main components of the two-proton
wave functions corresponding to the
state which in zeroth order is $(2s_{1/2})^2$ in Fig. \protect\ref{pp}
as a function of $G_\alpha$ ($\times 10^4$, in MeV). The
single-particle states are as in 
Table \protect\ref{spsp}.
Only components which in absolute value are larger than 0.2 are given.
\label{pps1}}
\begin{tabular}{cccccccc}
$G_\alpha$&$(1d_{5/2})^2$&$(2s_{1/2})^2$&$(0g_{7/2})^2$&$(1d_{3/2})^2$
&$(0h_{11/2})^2$\\
2 &------&(0.99,0.00)&------&------&------\\
6&(-0.28,0.00)&(0.67,0.00)&(0.61,-0.00)&(0.23,-0.00)&------\\
10&(-0.58,-0.00)&(0.42,-0.00)&(0.60,0.00)&(0.26,-0.00)&(-0.22,0.00)\\
12&(-0.68,-0.00)&(0.35,-0.00)&(0.54,0.00)&(0.25,-0.00)&(-0.22,0.00)\\
14&(-0.74,-0.00)&(0.31,-0.00)&(0.50,0.00)&(0.23,-0.00)&(-0.21,0.00)\\
16&(-0.78,-0.00)&(0.28,-0.00)&(0.46,0.00)&(0.22,-0.00)&(-0.20,0.00)\\
\end{tabular}
\end{table}
\begin{table}
\caption{Main components of the two-proton
wave functions corresponding to the
state which in zeroth order is $(1d_{5/2})^2$ in Fig. \protect\ref{pp}
as a function of $G_\alpha$ ($\times 10^4$, in MeV). The
single-particle states are as in 
Table \protect\ref{spsp}.
Only components which in absolute value are larger than 0.2 are given.
\label{ppd5}}
\begin{tabular}{cccccccc}
$G_\alpha$&$(1d_{5/2})^2$&$(2s_{1/2})^2$&$(0g_{7/2})^2$&$(1d_{3/2})^2$
&$(0h_{11/2})^2$&$(0i_{13/2})^2$\\
2 &(1.00,0.00)&------&------&------&------&------\\
6&(0.95,0.00)&------&(0.23,0.00)&------&------&------\\
10&(0.80,0.00)&------&(0.40,0.00)&(0.22,-0.00)
&(-0.29,-0.00)&------\\
12&(0.72,0.00)&(0.20,-0.00)&(0.45,0.00)&(0.25,-0.00)
&(-0.35,-0.00)&------\\
14&(0.65,0.00)&(0.21,-0.00)&(0.47,0.00)&(0.27,-0.00)
&(-0.40,-0.00)&(0.20,-0.01)\\
16&(0.60,0.00)&(0.21,-0.00)&(0.48,0.00)&(0.28,-0.00)
&(-0.43,-0.00)&(0.23,-0.01)\\
\end{tabular}
\end{table}
\begin{table}
\caption{Convergence of energies corresponding to the
states labelled $(1d_{5/2})^2$ in Fig. \protect\ref{pp} as a
function of the number of Gaussian points $n_g$.
The columns are labelled by the strength $G_\alpha$
($\times 10^4$, in MeV).
\label{snen}}
\begin{tabular}{cccccccc}
$n_g$&2&10&14&16\\
0&(4.996,0.000)&(3.118,0.142)&(1.150,0.349)&(-0.030,0.474)\\
10&(4.99275,0.00151)&(2.76320,0.16744)&(0.27597,0.41367)
&(-1.21486,0.56018)\\
50&(4.99316,0.00007)&(2.79745,0.00742)&(0.35026,0.01761)
&(-1.12083,0.02337)\\
100&(4.99302,-0.00000)&(2.79031,-0.00017)&(0.33314,-0.00040)
&(-1.14366,-0.00053)\\
150&(4.99309,-0.00000)&(2.79025,-0.00000)&(0.33299,-0.00000)
&(-1.14386,-0.00000)\\
200&(4.99309,-0.00000)&(2.79025,-0.00000)&(0.33299,-0.00000)
&(-1.14386,-0.00000)\\
\end{tabular}
\end{table}
\begin{table}
\caption{Two-proton wave function amplitudes corresponding to
$G_\alpha$ = 0.0016 MeV in Table \protect\ref{snen}
as a function of
the number of scattering states $n_g$ included in the basis.
Only components which in absolute value are larger than 0.2 are given.
\label{snwf}}
\begin{tabular}{cccccccc}
$n_g$&$(1d_{5/2})^2$&$(0g_{7/2})^2$&$(0h_{11/2})^2$&$(1d_{3/2})^2$
&$(0i_{13/2})^2$&$(2s_{1/2})^2$\\
0 &(0.641,0.025)&(0.478,0.000)&(-0.413,0.009)&(0.274,-0.003)
&(0.212,-0.022)&(0.208,-0.000)\\
10 &(0.592,0.023)&(0.479,0.003)&(-0.435,0.007)&(0.277,-0.002)
&(0.232,-0.023)&(0.206,0.000)\\
50 &(0.598,0.004)&(0.480,0.003)&(-0.433,-0.001)&(0.277,-0.000)
&(0.231,-0.014)&(0.206,0.000)\\
100 &(0.598,0.004)&(0.480,0.003)&(-0.433,-0.001)&(0.277,-0.000)
&(0.231,-0.014)&(0.206,0.000)\\
\end{tabular}
\end{table}

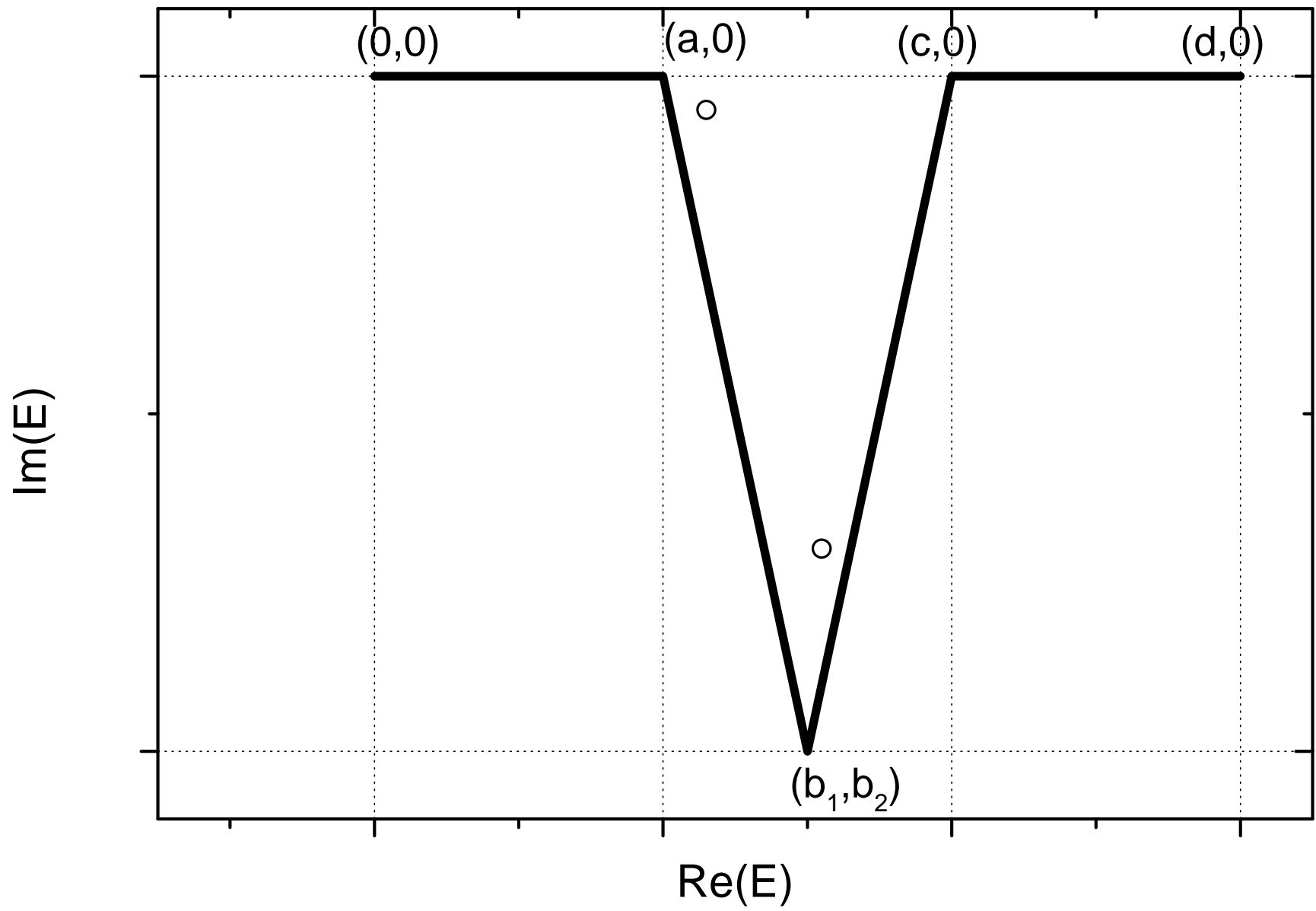
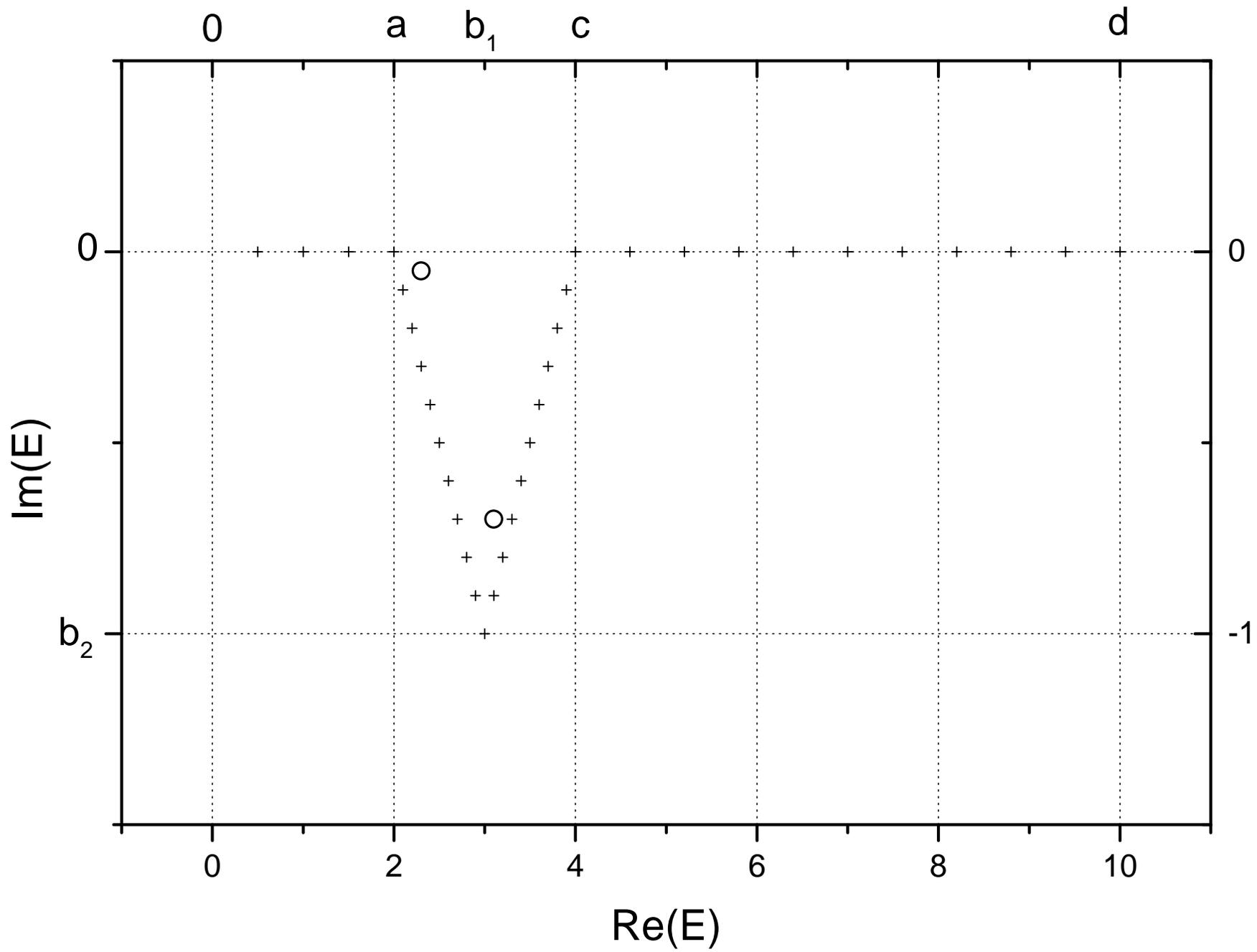
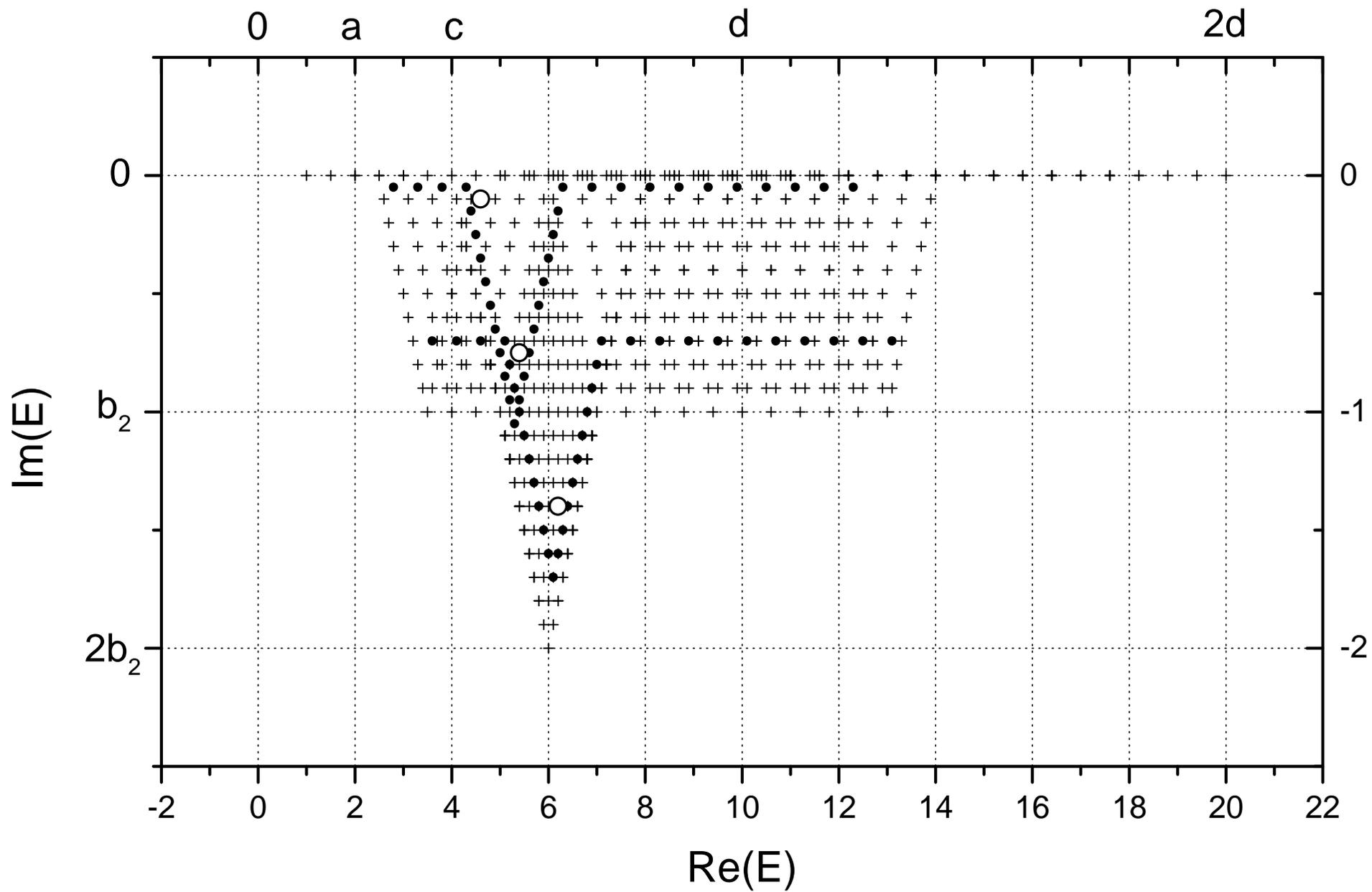
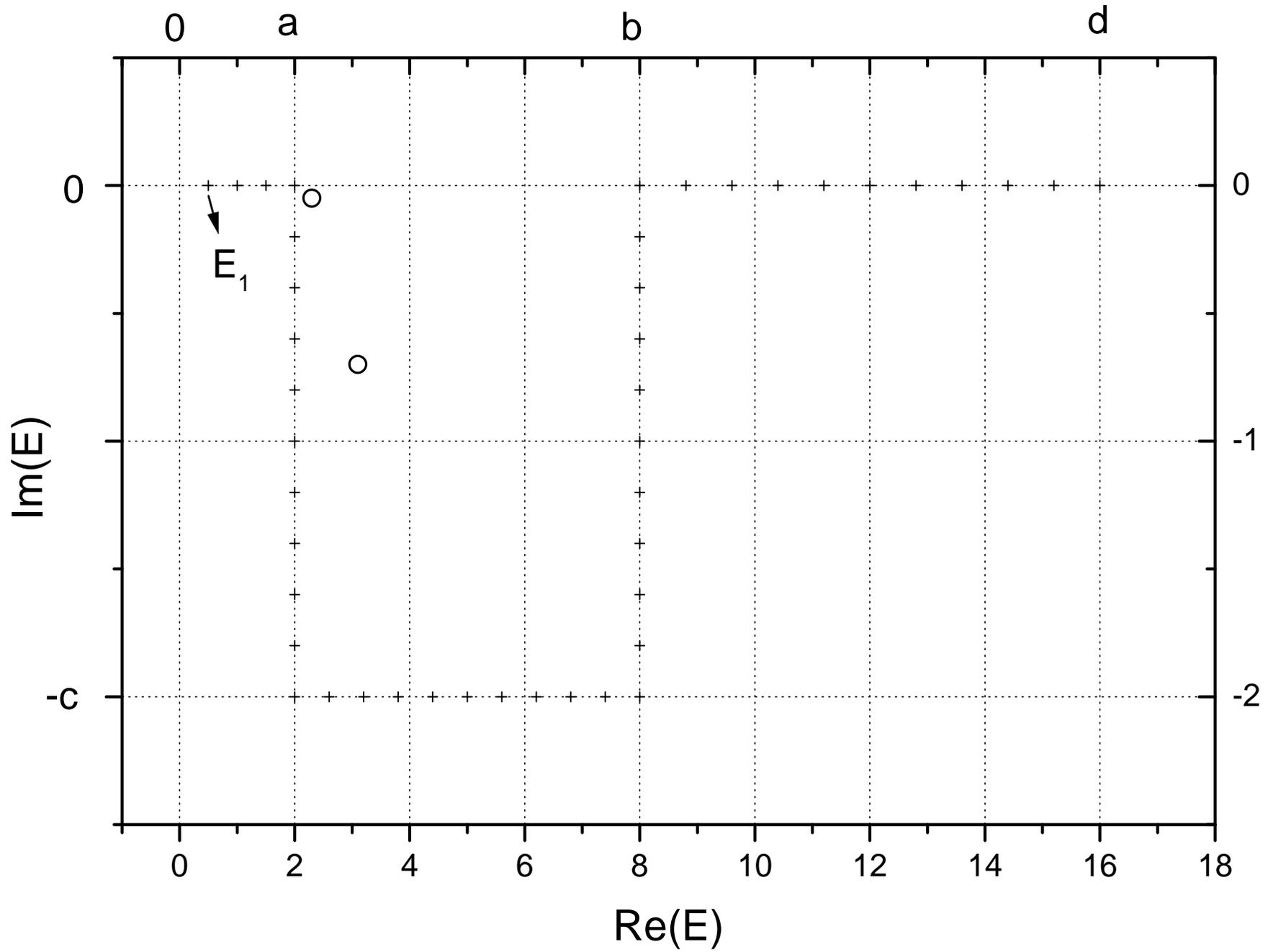
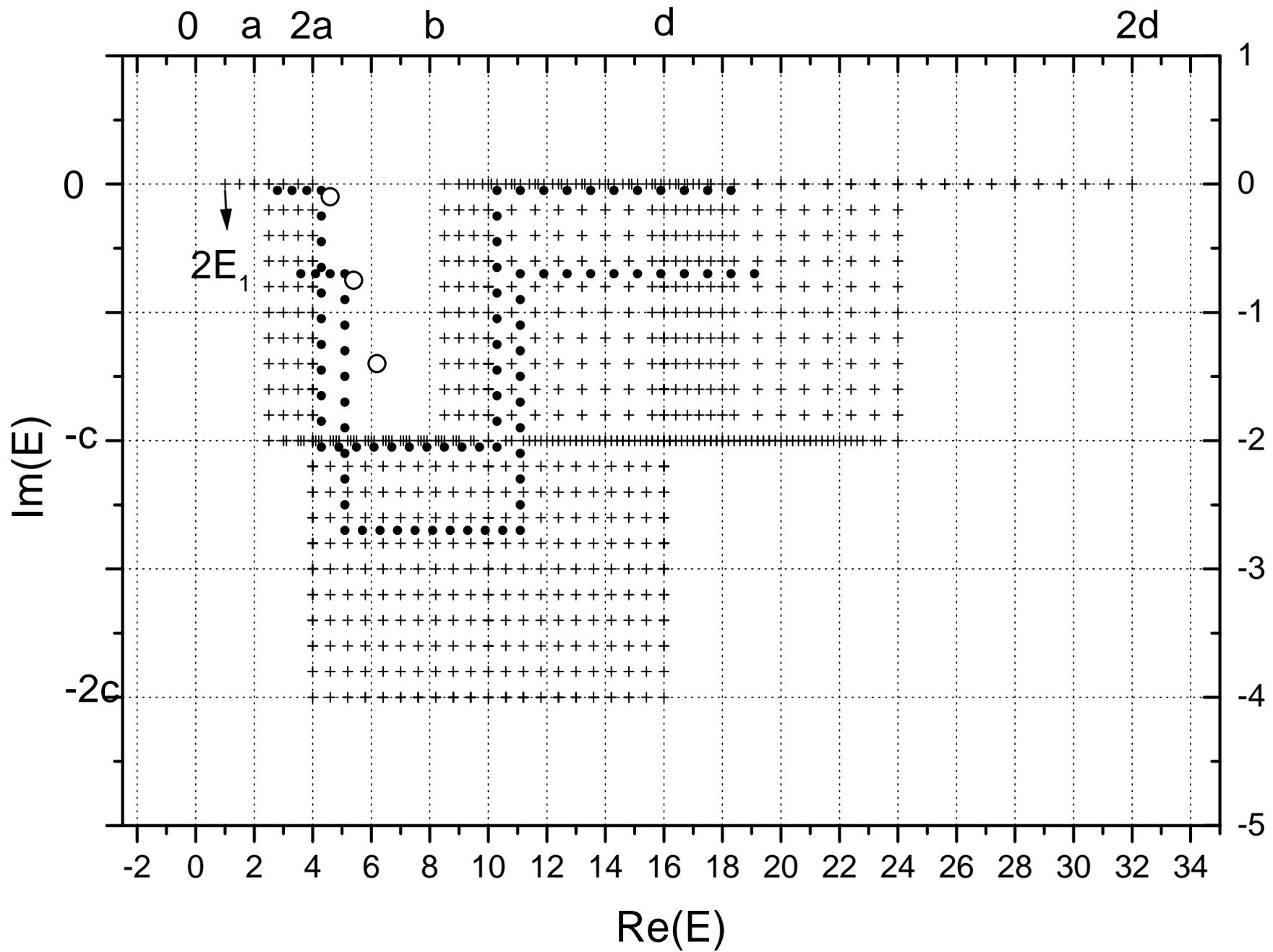
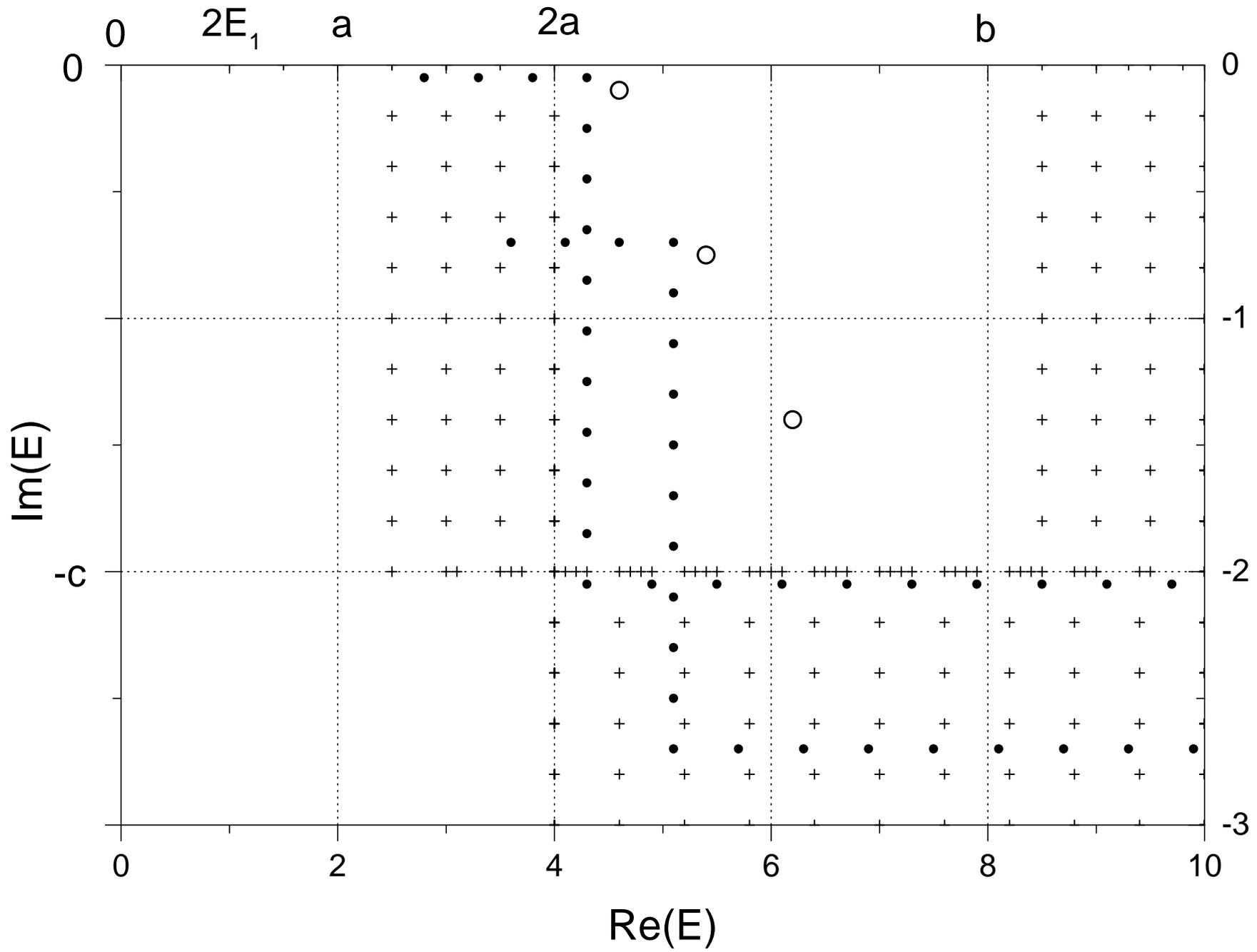
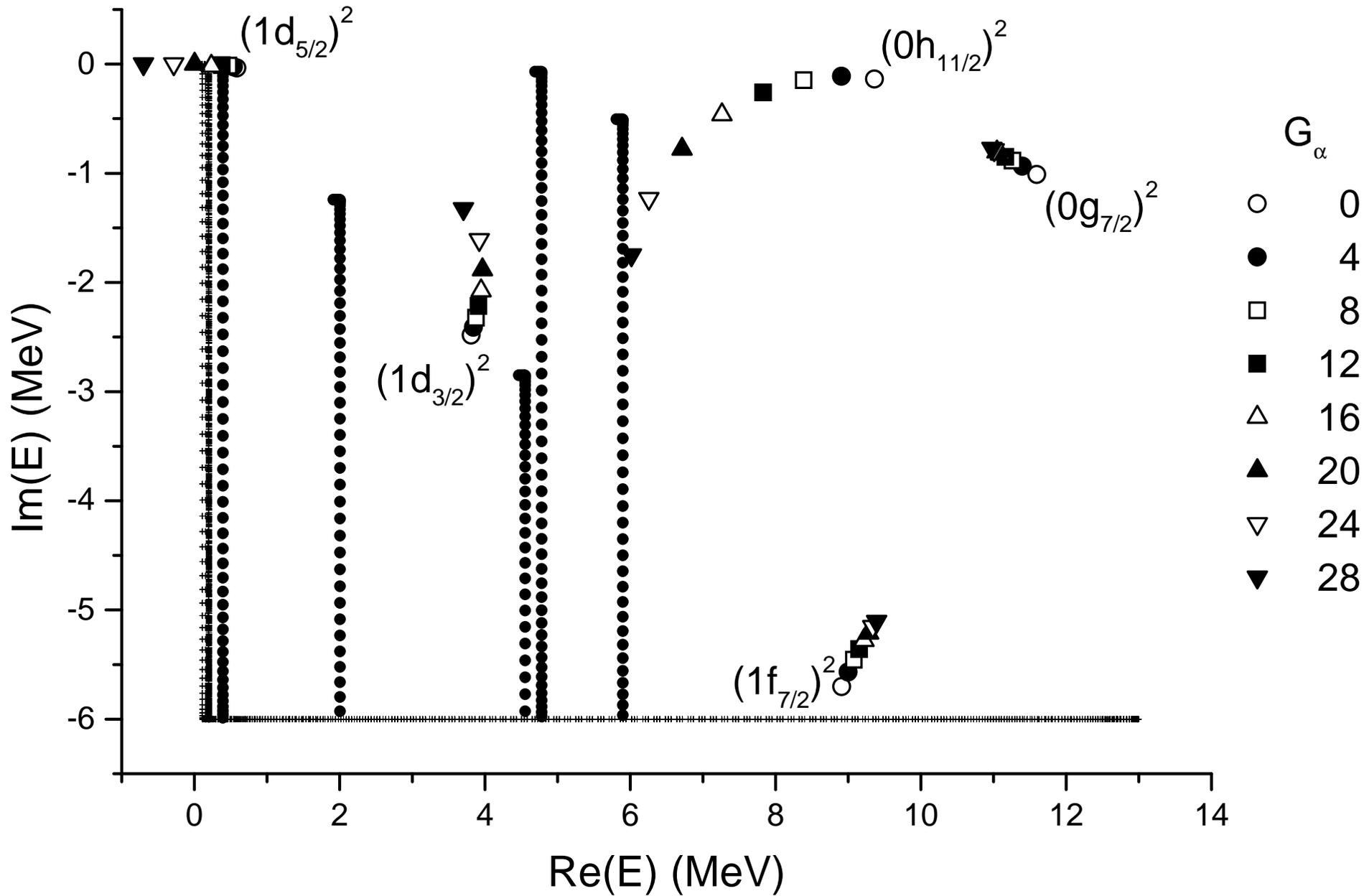
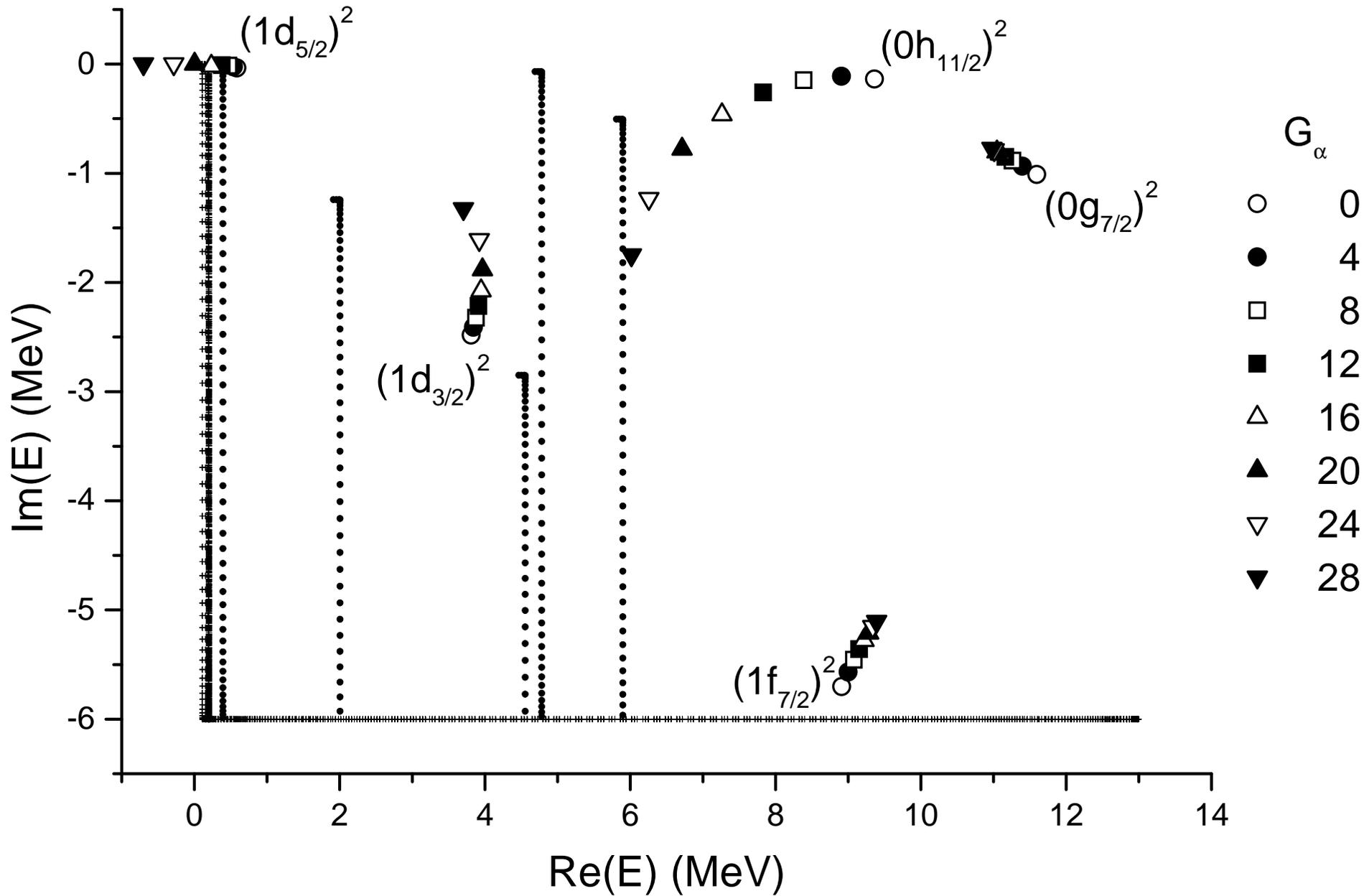
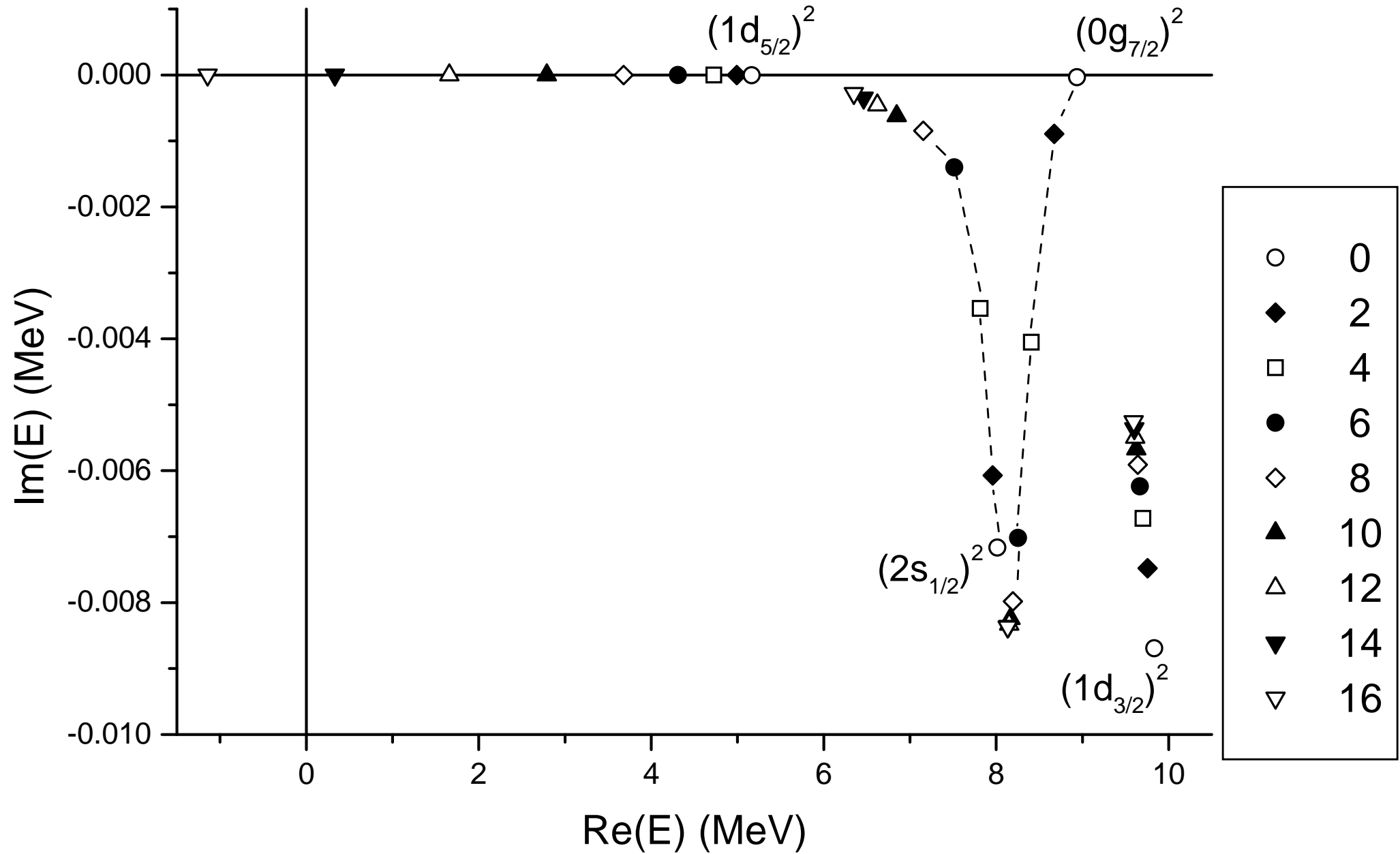
\end{document}